\documentclass[reprint]{revtex4-1}
\usepackage{graphicx}   

\begin{document}

\title{Jump at the onset of saltation}
\author{M. V. Carneiro$^1$}
\author{T. P\"ahtz$^1$  }
\author{H. J. Herrmann$^{1,2}$}
\affiliation{$^1$Institut f\"ur Baustoffe, ETH-H\"onggerberg, Schafmattstrasse 6, 8093 Z\"urich, Switzerland}
\affiliation{$^2$Departamento de F\'isica, Universidade Federal do Cear\'a, 60451-970 Fortaleza, Cear\'a, Brazil}
\date{\today}

\begin{abstract}
We reveal that the transition in the saturated flux for aeolian saltation is generically discontinuous by explicitly simulating particle motion in turbulent flow. This is the first time that a jump in the saturated flux has been observed. The discontinuity is followed by a coexistence interval with two metastable solutions. The modification of the wind profile due to momentum exchange exhibits a maximum at high shear strength.
\end{abstract}

\maketitle
Aeolian saltation is one of the main actors of molding Earth's landscape. 
It not only has impact on the evolution of the agricultural areas but also on the change of river courses and the motion of sand dunes. 
Saltation consists in the transport of sand or gravel by air or water and occurs when sand grains are lifted and accelerated by the fluid, hopping over the surface and ejecting other particles \cite{Greeley_book, Shao_Book}. The proper measurement of saltation close to the impact threshold through experiments is strongly limited by technical difficulties. Because of temporal fluctuations, one obtains unreliable data and large error bars. The computational simulation of aeolian saltation, however, can monitor every mechanical interaction between particles giving local insight about the particle splash and, in particular, about the system close to the impact threshold. Here, we report for the first time the existence of a discontinuity in the flux at the onset of saltation using discrete elements.

Saltation was first described by Bagnold's seminal work discussing the particle splash and proposing a cubic relation between wind shear velocity and saturated mass flux \cite{Bagnold1, Bagnold2, Bagnold3}. Greeley et al. \cite{Greeley} confirmed experimentally Bagnold's result and  Ungar and Haff \cite{Ungar} complemented it in their numerical model with a quadratic relation  near the impact threshold. Anderson and Haff \cite{Anderson, Anderson2} studied numerically the statistical properties of the particle splash relating the number and velocities of ejected particles to the impacting ones. Splash entrainment was carefully studied in experiments \cite{Rioual, Rioual_thesis, Rioual_2003, Oger} and used in nonsteady saltation models \cite{Sauermann, Parteli} finding that the splash mechanism dominates the saturation of sand flux. Recently, Almeida et al \cite {Almeida, Almeida_PNAS} implemented the full feedback with the fluid and fixed splash angle while Kok and Renno \cite{Kok} used a splash entrainment function, both resulting in good agreement with experiments. These and other theoretical studies, e.g., \cite{Andreotti, jenkins, McEwan}, however, describe the sand bed as a rough wall instead of resolving it at the particle scale and consequently rely on an empirical splash function. 

We present a discrete element simulation for aeolian saltation  which does not require the use of a splash function. Particles are subjected to gravity $g$ and dragged by a height-dependent wind field. The unperturbed wind profile is \cite{Shao_Book}

\begin{figure}[h]
{\includegraphics[scale=0.45]{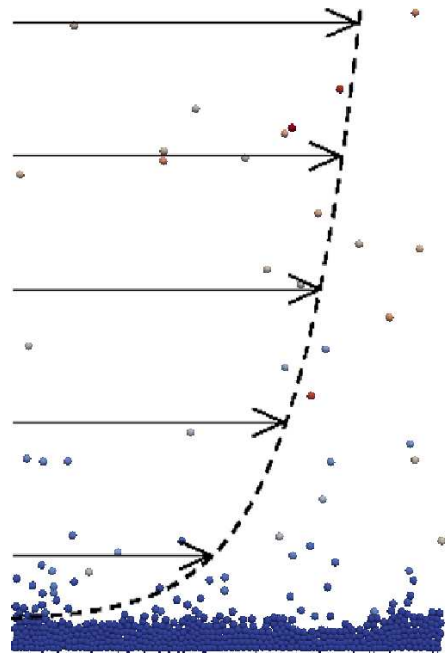}} 
\caption{[Color online] Snapshot of the simulation with an initially logarithmic wind profile. Colors represent the particle velocity.}
\label{sand_bed}
\end{figure}

\begin{equation}
u(y) = \frac{u_*}{\kappa}\ln\frac{y - h_0}{y_0},
\label{wind_profile_eq}
\end{equation}

\noindent where $y_0$ is the roughness of the bed, $h_0$ is the bed height, $\kappa = 0.4$ the von K\'arm\'an constant, and $u_*$ the wind shear velocity. Below $h_0$, the wind velocity is zero.
We adopted the widely used roughness law, $y_0 = D_{mean}/30$, measured by Refs. \cite{Nikuradse, Keulegan} for hydrodynamically rough pipe flow and confirmed by Ref \cite{Bagnold3} for air flow, where $D_{mean}$ is the mean diameter of the particles. A particle of diameter $D$ is accelerated by the wind drag force given by \cite{Shao_Book}  

\begin{equation}
F_d = -\frac{\pi D^2}{8} \rho_a C_d v_r \mathbf{v_r},
\label{drag_forces_eq}
\end{equation}

\noindent where $\rho_a$ is the air density and $\mathbf{v_r} = \mathbf{v} - \mathbf{u}$ is the velocity difference between particle and air, with $v_r = |\mathbf{v_r}|$.
The drag coefficient $C_d$ proposed by Cheng \cite{Cheng} is suited to model natural and irregularly shaped grains: 
\begin{equation}
C_d = \left[\left(\frac{32}{Re} \right)^{2/3} + 1 \right]^{3/2}, Re = \frac{\rho_a v_r D_{mean}}{\mu},
\label{drag_coefficient}
\end{equation}

\noindent where $\mu = 1.8702 \times 10^{-5} kg/(m.s)$ is the dynamic viscosity and $Re$ is the Reynolds number, which together with the Shields number $\theta$ below are the pertinent dimensionless parameters of our problem:

\begin {equation}
\theta = \frac{u_*^2}{(s-1) g D_{mean}} 
\end{equation}

\noindent where $s = \rho_s/\rho_w$ is the ratio between the grain and fluid density.

The detailed integration of the wind profile considering the momentum exchange is explained in the \emph{Supplemental Material} \cite{Sup}.

An aerodynamic lift arises from shear in the flow, which results in a pressure gradient normal to the shear in the direction of decreasing velocity.
This aerodynamic lift can be approximately described by \cite{Shao_Book} 

\begin{equation}
F_l = \frac{\pi D^3}{8}  \rho_a C_l \nabla v_r^2
\label{lift_forces}
\end{equation}

\noindent where the lift coefficient $C_l$ is proportional to $C_d$ \cite{Chepil}. 

The vertical motion of the particle is given by the competition between the gravity $g$, lift forces, and the rebouncing of particles with the ground. 

Alternatively, instead of lift forces, we also did some simulations perturbating the system at rest by lifting up at every second a fraction $c=0.2$ of the surface particles by a height $D_{mean}$ with a  probability $c = 0.2$ in order to restart the saltation.

We consider a disordered particle bed with $500$ spherical particles initially at rest in a two dimensional system. The diameters are randomly chosen from a Gaussian distribution around $D_{mean}$ of width $0.15 D_{mean}$. We simulated systems with more particles to verify that the bottom wall effects are negligible as discussed in Ref \cite{Rioual, Rioual_thesis}. Within the error bars, they displayed identical properties and therefore we did not need to consider larger systems. At $t=0$, some particles are dropped from randomly chosen heights and when they reach the ground they collide with particles at rest inside the bed and thereby trigger saltation. The particle collisions are computed using the discrete elements method.

Trajectories are obtained by integrating the equations of motion according to the velocity-St\"ormer-Verlet scheme \cite{Griebel}, using a spring dashpot potential with the spring constant $k$ and a dissipative damping parameter $\gamma$. Further details about the technique are explained in the \emph{Supplemental Material} \cite{Sup}. The dissipative rate of the lower boundary was set high enough, $\gamma_w = 0.8$, to avoid that the shock wave generated by the impact reaches the system boundaries and be reflected to the bed surface, ejecting additional grains \cite{Rioual_thesis}.

Space is sliced in vertical rectangular domains of size $(250 \times 75) D_{mean}$. The top is placed sufficiently high to mimic an open system. Periodic boundary conditions are imposed in wind direction and top and bottom boundaries are reflective. 
 \begin{figure}[h]  
\rotatebox{-90}{{\includegraphics[scale=0.30]{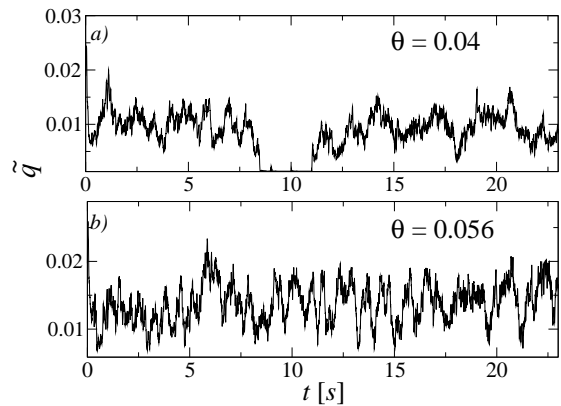} }}
\caption{Flux series for two different Shields numbers near the onset. In $(a)$, the system was perturbed at $t = 9, 10$, and $11s$ to keep the flux different from zero. The small peaks at t = 9 and 10s show that some particles were lifted but did not provoke a splash. In $(b)$ the system never settled down.}
\label{flux2}
\end{figure}

We measure the particle flux through
 
\begin{equation}
q = \frac{1}{A}\displaystyle\sum\limits_{i}^N m_i v_{i}^{x}
\label{saturated_flux}
\end{equation}

\noindent where $v_{i}^{x}$ is the particle velocity in the $x$ direction. The saturated flux is the average flux in the stationary state. Simulated systems with different number of particles display similar flux.
The dimensionless flux can be obtained by
 
\begin{equation}
\tilde{q}  = \frac{q}{\rho_s\sqrt{(s-1)gD^3_{mean}}}
\label{dimensionless_flux}
\end{equation}
\noindent where $\rho_s$ is the grain density.

\begin{figure}
\begin{center}
\rotatebox{-90}{\includegraphics[scale=0.37]{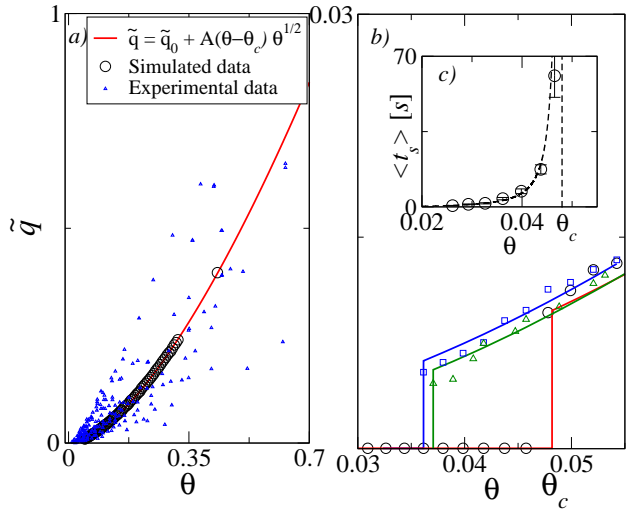}}
\caption{[Color online] Saturated flux as function of the Shieds number. $(a)$ Simulated data fitted by $\tilde{q}_s = \tilde{q}_0 + A (\theta - \theta_c)\theta^{1/2}$, $\tilde{q}_0 = 0.01$, $A = 1.52$, and $\theta_c = 0.048$ in comparison with experimental data \cite{Iversen}.$(b)$ Detailed view of the metastable region of the discontinuous transition. The green curve with triangles and blue lines with squares are the results including perturbation with $c =0.2$ and lift forces with $C_l = 0.425 C_d$, respectively. $(c)$ Average transient time to settle without perturbation in the metastable region. The dashed line fits the time distribution according to $t_s = t_0 [\theta/(\theta_c - \theta)]^{p}$ with $t_0 = 0.524 s, p = 1.5, \theta_c = 0.048 $.}
\label{diagram_fg}
\end{center}
\end{figure} 

\begin{figure}[h]
\rotatebox{-90}{\includegraphics[scale=0.35]{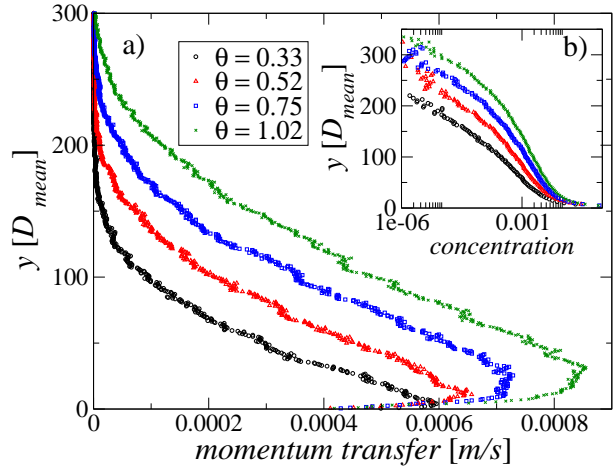}}
\caption{[Color online] Momentum exchange in $y$ direction becomes nonmonotonic as function of height as we increase the Shield number. In the bottom part, it increases with the wind velocity due to the high concentration of particles. The concentration shown in the inset decays logarithmically for lower velocities but has a pronounced inflection point for higher velocities.}
\label{momentum}
\end{figure}

We verified numerically that the dimensionless saturated flux remains invariant under changes of the parameters as long as the Reynolds number and Shields number are fixed.

Simulations verify the existence of an onset velocity for sustained flux $\theta_c = 0.048$ with strong temporal fluctuations as illustrated in Fig. \ref{flux2}.
Below this value the flux will stop after some time.
This happens because an impact may not necessarily eject other particles that can carry on saltation, and
once the flux is zero it cannot restart unless a perturbation or lift force is introduced.

In Fig.\ref{flux2}(a) below the onset, some perturbations were introduced at $t = 9, 10$, and $11s$ in order to trigger saltation, after it stopped. 
The perturbation at $t = 11s$ restarted the saltation and the flux returned to the previous levels. 
For $ \theta_t < \theta < \theta_c$, the system displays this kind of metastable behavior with two possible solutions, either saltation or no motion, strongly dependent on the initial conditions and the triggering mechanism. 
$\theta_t$ is the Shields number of the threshold, below which no particle is transported and $\theta_c$ is a critical Shields number separating the metastable region from the normal regime of saltation.
The previously defined perturbations or lift forces according to Eq. \ref{lift_forces} are not sufficient to restart saltation for $\theta < \theta_t$, where $\theta_t \simeq 0.037$ for perturbations with $c=0.2$ and $\theta_t \simeq 0.036$ for lift forces with $C_l = 0.425C_d$. The lower bound of the metastable region changes depending on the perturbation probability or the lift constant.   They are no longer needed to maintain the saltation for $\theta \ge \theta_c$. The fact that perturbations are necessary to sustain saltation underlines the importance of the turbulent lift forces in the metastable region.

Fig. \ref{diagram_fg}(a) presents the saturated flux for different Shields numbers in comparison with field and wind tunnel experimental data from Iversen and Rasmussen \cite{Iversen}.
The calculated data are well fitted using $\tilde{q}_s = \tilde{q}_0 + A (\theta - \theta_c)\theta^{1/2}$ with $A = 1.52$ and $\tilde{q}_0 = 0.01$, which is similar to the relation proposed in Ref.\cite{lettau}. Figure \ref{diagram_fg}(b) presents details of the discontinuous transition at $\theta_c$ with a jump $\tilde{q}_0$ in the saturated flux. 
The green triangles are the data including perturbations and the blue squares are the data with lift forces.
Additional simulations show that $\theta_t$ and the jump $\tilde{q}_0$ depend on the lift coefficient $C_l$.

Figure \ref{diagram_fg}(c) presents the average transient time $<t_s>$ it takes for the system to settle without perturbation or lift diverging when $\theta \rightarrow \theta_c$. The dashed line fits the data according to $t_s = t_0 [\theta/(\theta_c - \theta)]^{p}$ with $t_0 = 0.052 s, p = 1.5$.
 The transient times $<t_s>$ were averaged over $60$ simulations with different initial conditions. 

The discontinuity in the saturated flux is verified also at the same value $\theta_c$ using the drag coefficient from Ref. \cite{Durst} and is shifted in the velocity axis by the dissipation rate $\gamma$, i.e., to lower (higher) critical velocities for lower (higher) dissipation rates.
Interestingly, we were able to find the same discontinuous transition and metastable region also using the code of Ref \cite{Kok}.  

\begin{figure}
\includegraphics[scale=0.28]{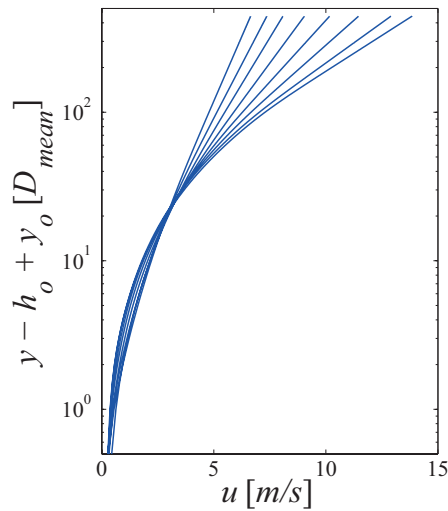} 
\caption{[Color online] Modified wind profiles intersect for $\theta > \theta_c$  at the Bagnold focus. This estimate corroborates measurements for a particle size $D_{mean}=200\mu m$ \cite{Andreotti, Ungar, Werner, Kok}. }
\label{bagnold_focus}
\end{figure}

Figure \ref{momentum} shows the momentum transfer to the particles as a function of height for different Shields numbers. Because of the higher concentration of particles, the largest momentum transfer occurs close to the bed surface. Additionally, this region coincides with the one in which we observe small changes in the modified wind profile in Fig. 1 in the \emph{Supplemental Material} \cite{Sup}. Momentum transfer for high Shields number exhibits a local maximum which gets more pronounced with increasing the Shields number. 
As shown in Fig. \ref{momentum}(b), high Shields numbers enhance the concentration at higher $y$ where the wind profile grows stronger. This increases momentum transfer at this level.
In Fig. \ref{momentum}(b), we see that in fact the concentration exhibits a saddle point at the same height at which the maximum in momentum transfer appears.

Figure \ref{bagnold_focus} shows modified velocity profiles for $\theta > \theta_c$ crossing each other at a focal point, called the "Bagnold focus" \cite{Bagnold3} and approximately located $0.5cm$ above the sand bed, which is in qualitative agreement with measurements and theory \cite{Andreotti, Ungar, Werner, Kok}. 

In summary, we performed particle simulations without any assumption concerning particle trajectories or splash and revealed that the transition is discontinuous with a metastable state for aeolian saltation never reported before. In the metastable region, perturbations or lift forces are required to keep saltation going. It would be interesting to reanalyze or remake experiments of saltation very close to the onset to verify the predicted jump. The existence of a discontinuity at the threshold of turbulent particle transport can have far-reaching consequences in numerous applications ranging from dune mitigation or pneumatic transport to the understanding of Martian landscapes.

This work has been supported by the National Council for Scientific and Technological Development CNPq, Brazil and ETH Grant (No. ETH-10 09-2). Authors also acknowledge discussions with Dirk Kadau, Allen Hunt and Klaus Kroy.

\bibliography{aeolian}
\end{document}


\title{Supplemental Material: Jump at the onset of saltation}
\author{M. V. Carneiro$^1$}
\author{T. P\"ahtz$^1$  }
\author{H. J. Herrmann$^{1,2}$}
\affiliation{$^1$Institut f\"ur Baustoffe, ETH-H\"onggerberg, Schafmattstrasse 6, 8093 Z\"urich, Switzerland}
\affiliation{$^2$Departamento de F\'isica, Universidade Federal do Cear\'a, 60451-970 Fortaleza, Cear\'a, Brazil}
\date{\today}

\maketitle

\section{The wind profile}

The description of the wind considers a logarithmic profile given by

\begin{equation}
u(y) = \frac{u_*}{\kappa}\ln\frac{y - h_0}{y_0}.
\label{wind_profile_eq}
\end{equation}

But when particles are accelerated also the wind is decelerated, modifying the wind profile \cite {Anderson2}. We obtain the grain stress profile $\tau_g(y)$ by integrating the drag forces acting on all particles above height $y$, as  

\begin{equation}
\tau_g(y) =  \int\limits_y^\infty f(y')dy' \cong \sum\limits_{j:Y^j > y} \frac{F^j(u_j)}{A},  
\label{grain_stress_profile}
\end{equation}

\noindent where $f(y')$ is the average force per unit volume and $F^j$ the drag force applied on the particles at a position $Y^j$ above $y$, which depends on the wind velocity $u_j$ at the same position, and $A$ is the cross-sectional area parallel to the ground assuming a finite system thickness. The grain stress profile is then used to calculate the modified wind profile by rewriting Eq. (\ref{wind_profile_eq}) as a differential equation,

\begin{equation}
\frac{du}{dy} = \frac{u_\tau(y)}{\kappa y},
\label{wind_profile_2_eq}
\end{equation}

\noindent where the modified wind strength  $u_\tau(y)$ depends according to Anderson and Haff \cite{Anderson} on the grain stress (Eq. \ref{grain_stress_profile}) as  

\begin{equation}
u_\tau(y) = u_* \sqrt{1-\frac{\tau_g}{\rho_a u^2_*}},
\label{wind_profile_3_eq}
\end{equation}

The numerical solution is achieved iteratively. If no particle is accelerated $u_\tau(y) = u_*$ and consequently, the unperturbed logarithmic profile is obtained. Otherwise, a grain stress arises reducing the shear velocity in the underneath areas.

The wind profile is recalculated at each iteration step starting from the top of the bed for which basis position $h_0$ and shape change dynamically. Consequently, we need to recalculate also $h_0$ each time.  
Highly concentrated areas like the one close to the sand bed reduce strongly wind velocities. If the calculated velocity $v_i$ in area $y_i$ is below $0.1u_*$, this area always contains bed particles and the velocity is set to zero. This means that $h_0$ is chosen to be the point where the calculated velocity exceeds $0.1u_*$.

\begin{figure}  
\includegraphics[scale=0.30]{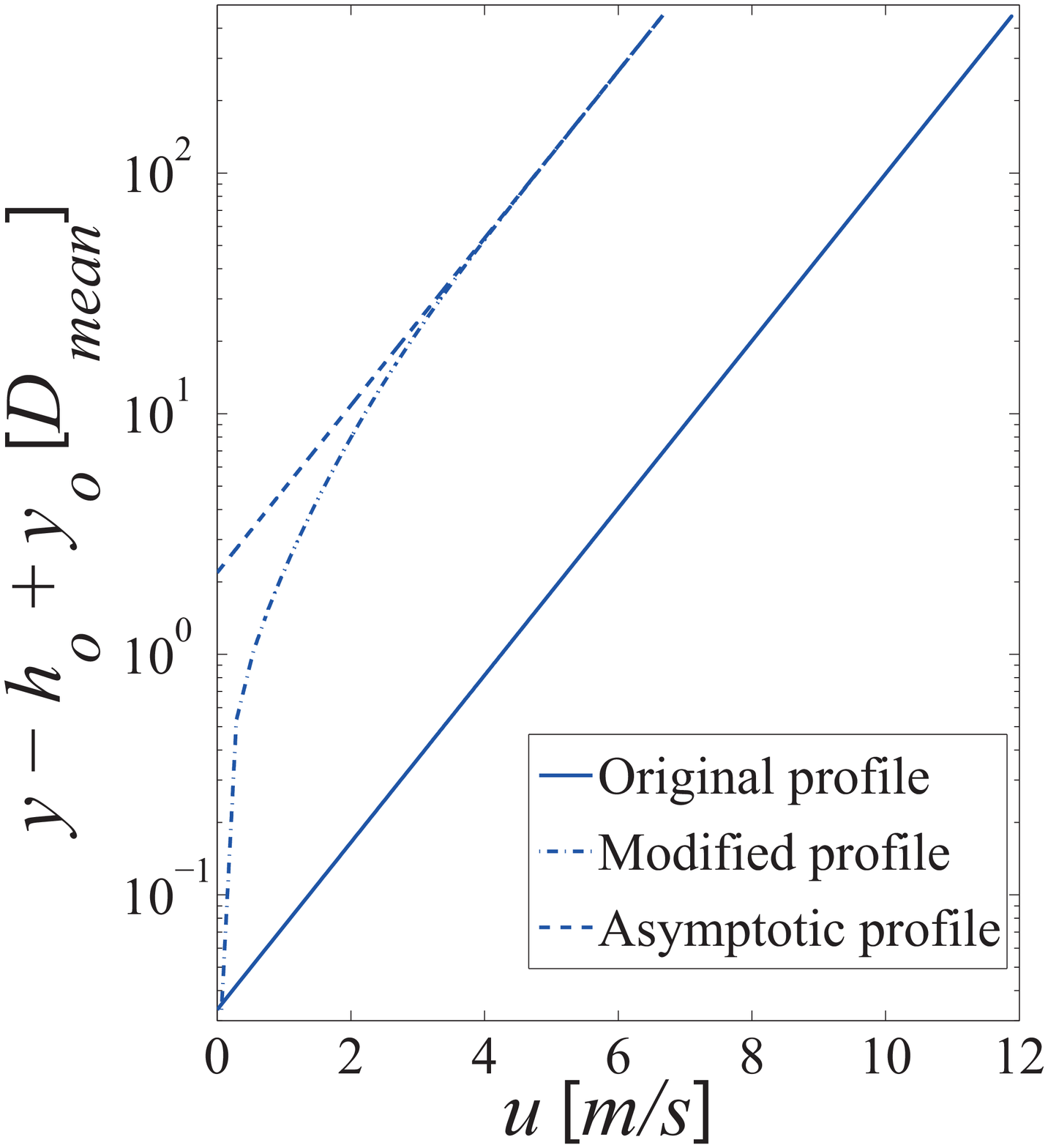} 
\caption{[Color online] Wind profiles with and without momentum exchange. The modified profile has a weaker increase at highly concentrated regions but an asymptotic slope equal to the unmodified one. 
At $y = h_0$, we have $u = 0$.}
\label{wind_profile}
\end{figure}

Figure \ref{wind_profile} shows an example of the wind profile for $u_* = 1.4$ $m/s$. As the wind strength decays with the number of saltating particles,  
the slope of the modified curve falls much below the original profile in the region nearby the particle bed. Because few or no particles are located in the higher regions, the slope of the modified curve  asymptotically converges to the unmodified one.

\section{Partice dynamics}

Molecular dynamics is used to solve numerically the Newton's equations of motion. 
The differential equations are iteratively solved using time-discretization obtaining the new positions and velocities of the particles from the old positions, old velocities, and the corresponding forces. An efficient  and, at the same time, stable approach for the time discretization of Newton's equations is the Verlet algorithm which builds on the integration method of St\"ormer found in Ref \cite{Griebel}. Let $\mathbf{x}_i^n$ , $\mathbf{v}_i^n$ and $\mathbf{F}_i^n$ be the vectors corresponding
to the position, velocity and force of the particle $i$ at time step $n$. Velocity and position of particle $i$ at the next time step can be obtained by \cite{Griebel}

\begin{eqnarray}
\mathbf{x}_i^{n+1} &=& \mathbf{x}^n_i + \delta t \mathbf{v}_i^n + \frac{\mathbf{F}^n_i \delta t^2}{2m_i}  \\
\mathbf{v}_i^{n+1} &=& \mathbf{v}^n_i + \frac{(\mathbf{F}_i^n + \mathbf{F}_i^{n+1})\delta t }{2m_i}
\end{eqnarray}  

\noindent where $\delta t$ and $m_i$ are the time step and the particle mass. $\mathbf{F}_i$ is the resultant force of all particles acting on particle $i$ \cite{Herrmann}

\begin{equation}
\mathbf{F}_i =  \sum^{N}_{j=1, j \neq i} \mathbf{F}_{ij}
\end{equation}

When two particles $i$ and $j$ overlap (i.e. when their distance is smaller than the sum of their radia) two forces act on the particle $i$, an elastic restoration force, 

\begin{equation}
F^{(i)}_{el} = k m_i[\vert \mathbf{r}_{ij} \vert - 1/2(d_i + d_j)] \frac{\mathbf{r}_{ij}}{\vert \mathbf{r}_{ij} \vert}
\end{equation}

\noindent where $k$ is a spring constant, $m_i$ is the mass of particle $i$, and $\mathbf{r}_{ij}$ points from particle $i$ to $j$ and a dissipation due to the inelasticity of the collision,

\begin{equation}
F^{(i)}_{diss} = -\gamma m_i(\mathbf v_{ij} \mathbf r_{ij}) \frac{\mathbf{r}_{ij}}{\vert \mathbf{r}_{ij} \vert^2}
\end{equation}

\noindent where $\mathbf{v}_{ij}= \mathbf{v}_i - \mathbf{v}_j$ is the relative velocity and $\gamma$ is phenomenological dissipation coefficient. In our work, we chose $k = 0.5$ and $\gamma = 0.3$ for particle collisions. The coefficient of restitution is given by the ratio between the velocities after and before the collision. We neglect Coulomb friction and rotation of particles. 
When a particle collides with a wall the same forces act as if it would have encountered another particle of diameter $d_0$ at the collision point \cite{Herrmann}.

\bibliography{supplement}